\documentclass[fleqn,usenatbib,onecolumn]{mnras}
\usepackage{newtxtext,newtxmath}
\usepackage[T1]{fontenc}
\usepackage{amsmath, amsfonts, graphicx, chngcntr, multirow, enumerate, float, hyperref}
\usepackage[dvipsnames]{xcolor}
\hypersetup{
	colorlinks=true,            
	urlcolor=MidnightBlue,      
	linkcolor=WildStrawberry,   
	citecolor=ForestGreen,      
}
\DeclareRobustCommand{\VAN}[3]{#2}
\let\VANthebibliography\thebibliography
\def\thebibliography{\DeclareRobustCommand{\VAN}[3]{##3}\VANthebibliography}

\newcommand{\luna}{{\texttt luna}}
\newcommand{\python}{{\texttt Python}}
\graphicspath{{./}{figures/}}

\title[Binary Planet Transits]{Transit Duration and Timing  Variations from Binary Planets}

\author[Chakraborty \& Kipping]{
Joheen Chakraborty,$^{1,2}$\thanks{E-mail: joheen@mit.edu}
David Kipping,$^{2}$
\\
$^{1}$MIT Kavli Institute for Astrophysics and Space Research, Massachusetts Institute of Technology, Cambridge, MA 02139\\
$^{2}$Department of Astronomy, Columbia University, 550 W 120th Street, New York, NY 10027\\
}

\date{Accepted 2022 December 1. Received 2022 November 21; in original form 2022 September 16}

\pubyear{2022}

\begin{document}
\label{firstpage}
\pagerange{\pageref{firstpage}--\pageref{lastpage}}
\maketitle

\begin{abstract}
Systems of two gravitationally bound exoplanets orbiting a common barycenter outside their physical radii (``binary planets") may result from tidal capture during planet-planet scattering. These objects are expected to form in tight orbits of just a few times their summed radii due to dynamical tides. As a result of their close proximity, their transits overlap heavily, leading to the deceptive illusion of a single planet of larger effective size, an effect compounded in the presence of noisy data and/or long integration times. We show that these illusory single-component transits, dubbed ``chimera transits", exhibit large-amplitude Transit Duration Variation (TDV) effects on the order of hours, as well as smaller Transit Timing Variations (TTVs). We compute an analytic approximation for the transit duration upper bound, assuming binary planets with low impact parameter and orbits coplanar with the stellarcentric orbit. We verify the accuracy of our expressions against dynamical simulations of binary Jupiters using the \luna\ algorithm, and provide a \python\ code for numerical calculations of the TDV signal in binary planet systems (\href{https://github.com/joheenc/binary-planet-transits}{github.com/joheenc/binary-planet-transits}). Additionally, chimera transits from binary planets exhibit TTVs of detectable amplitude and high frequency, falling within the recently identified exomoon corridor. Due to their anomalous shapes, depths, and durations, such objects may be flagged as false positives, but could be clearly surveyed for in existing archives.
\end{abstract}


\section{Introduction} \label{sec:intro}
Astronomical objects across a wide range of mass scales---from asteroids to stars to black holes---can exist in binary systems. Studying the ubiquity and characteristics of these binary populations is key to our understanding of dynamics and formation mechanisms across many physical regimes. However, although over 5100 exoplanets have been discovered to date\footnote[1]{\href{https://exoplanetarchive.ipac.caltech.edu/}{https://exoplanetarchive.ipac.caltech.edu/}} \citep{akeson13}, the case of a ``binary planet"---defined as a system of two planets orbiting a common barycenter outside their physical radii (e.g. the Pluto-Charon system, as opposed to an exomoon where the barycenter lies within the planet, \cite{sartoretti99,stern02})---has eluded us.

This non-detection is not due to a definitive consensus that binary planets cannot exist: in fact, there are multiple formation mechanisms discussed in the literature which could possibly give rise to systems of binary compact planets. Though generally thought that in-situ formation of similar-mass planetary companions within a protoplanetary disk is infeasible \citep{canup06}, it has been shown that in-situ formation followed by tidal evolution \citep{moraes20}, or even independent formation followed by tidal capture \citep{hamers18}, is a possible explanation for massive exomoons. On the scale of binary super-Earths, \cite{chrenko18} found that close encounters of multiple super-Earth embryos assisted by disk gravity can result in stable binary planets with a $\sim 10^4$ year lifetime before merger into the core of a giant planet. On the scale of binary Jupiters, the most promising formation channel arises from planet-planet scattering, an important dynamical mechanism thought to play a key role early in the formation and evolution of many planetary systems \citep{rasio96, ford08}. Though these events typically result in the ejection of planets from the system or collision of the planets, planet-planet tides from orbital crossings of two gas giants may result in tidal capture and circularization into a binary planet system in about 5-20\% of cases, loosely increasing with stellarcentric semimajor axis of the initial interaction \citep{podsiadlowski10}. \cite{ochiai14} found similar results for initial three-body interactions, finding that tidal dissipation resulted in binary planets in $\sim$10\% of cases from 0.5-10 AU, almost independent of semimajor axis. Subsequent long-term evolution occurs due to quasi-static planet-planet and planet-star tides, resulting in typical eccentricities of zero and separations of just a few times their summed radii. Furthermore, for orbital barycentric radii $\geq 0.3$ AU, the binaries are not destroyed during the main-sequence lifetime of a Sun-like star \citep{ochiai14}.

There has also been working looking into long-term stability of massive satellites of exoplanets. To first order, we expect a requirement that binary planets or exomoons should exist between the Roche and Hill spheres of the primary. In fact multiple studies \citep{domingos06,rosario20,quarles21} find that planet-satellite systems are only stable for separations within about \textit{half} the combined planetary Hill sphere, with a dependence on the orbital eccentricity of the planet, aligned qualitatively with the findings of \cite{ochiai14} that only compact binary orbits remain stable and that orbit tend to circularize. \cite{barnes02} found that giant planets cannot host massive satellites at close stellarcentric orbits, but by about 0.3 AU a Jupiter-like satellite can host an Earth-like satellite, and that no meaningful mass limits can be placed on satellite mass beyond $\sim 0.6$ AU. Notably, \cite{spalding16} found that in planet-moon systems which experience a significant inward migration event, evection resonance can remove any companions within about $\sim$10 planetary radii of the host---this result is in direct tension with the findings of existing binary planet stability models \citep{podsiadlowski10,ochiai14,lewis15}, and continued non-detection of binary planets may point to this mechanism as an underlying cause.

\begin{figure}
    \centering
    \includegraphics[width=\textwidth,keepaspectratio]{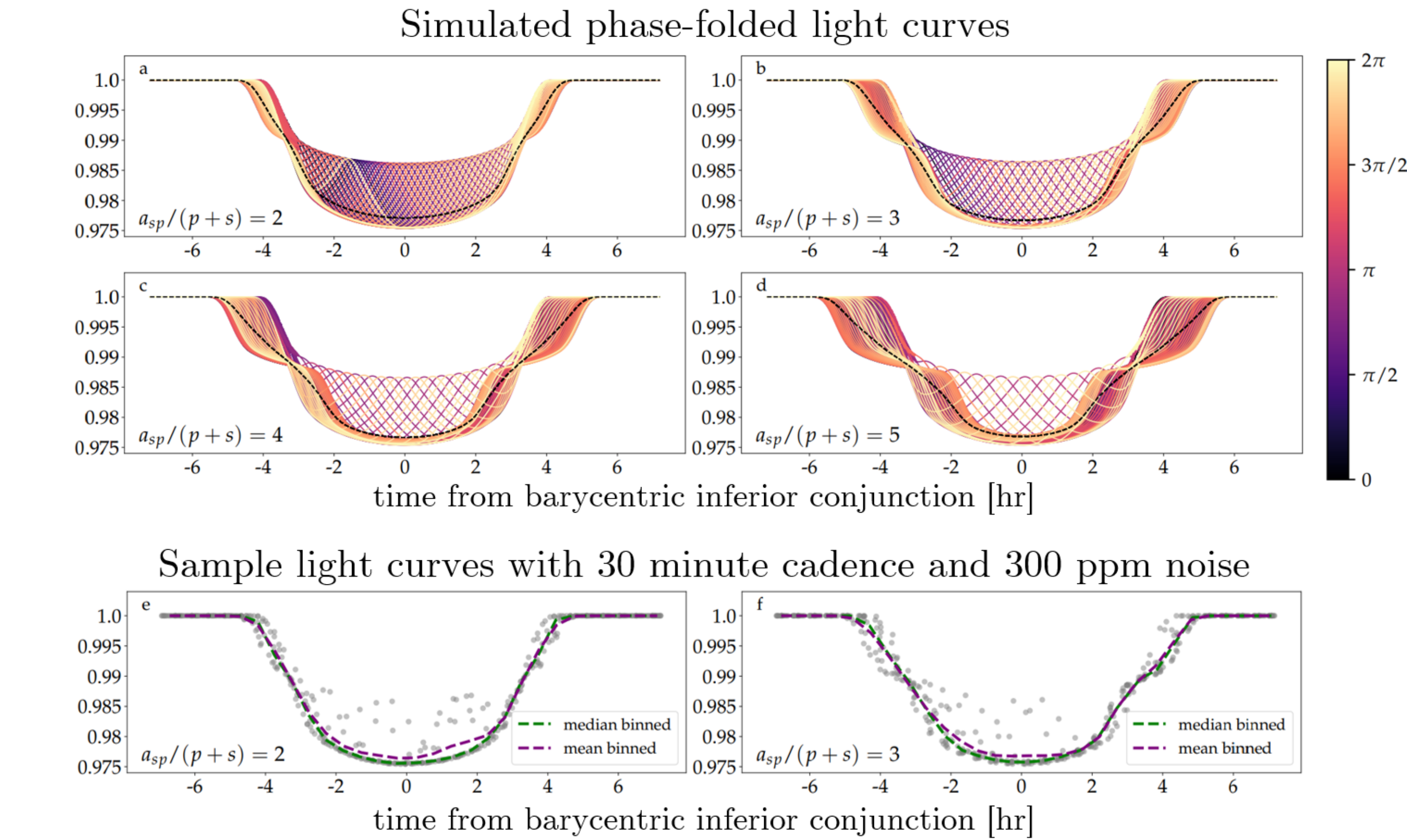}
    \caption{\textbf{a-d:} Transit light curves of a binary Jupiter around a Sun-like star calculated via numerical simulations. Due to the small orbital separation required for formation and stability of binary planet systems, the transits of both planets overlap and combine to create a single signal, the ``chimera transit". \textbf{e-f:} We also show simulated ``realistic" light curves of 20 phase-folded transit observations with a 30 minute cadence and 300 ppm injected noise. These light curves are not easily distinguished from single-planet transits, which may explain the lack of binary planets discovered from direct light curve inspection thus far.}
    \label{fig:transit}
\end{figure}

After formation and stability considerations, the question then becomes whether these systems leave any detectable signatures distinguishing them from a single-planet fit. Initially, \cite{podsiadlowski10} found that deviations from conventional radial velocity curves in binary Jupiter systems would be of order 0.3 cm s$^{-1}$, too small to be detected in current observations. More recent work in \cite{vanderburg18} and \cite{vanderburg21} found that precise constraints on binary planets can be placed from radial velocities on directly imaged exoplanets themselves; this is an exciting direction of further study considering the advances in directly imaged systems and high-resolution spectroscopy, but is not our focus in this paper. \cite{agol15} suggested the idea of spectroastrometry for detection of exomoons; however, in the case of binary planets where both bodies are presumed to have similar physical properties, and thus similar spectra, this technique becomes difficult. \cite{lewis15} further noted the difficulty of detection via gravitational microlensing (such events would be too infrequent) or the Rossiter-McLaughlin effect (the sample of sufficiently bright stars with transiting gas giants beyond 0.3 AU was too limited at the time of writing; however, we now note the presence of a sample of post-Kepler exoplanets beyond $a > 0.3$ AU and $m_V < 10$ which may provide an attractive sample for this approach, e.g. \cite{delrez21}). Instead, \cite{lewis15} turned to measurement of modulations of the transit light curve from a single-planet fit and deviations from transit to transit. However, given that these systems only remain stable for small separations of the individual planets, the transits will appear to be heavily intermixed. Moreover, the orbital period of the binary will not in general be in resonance with the stellarcentric orbital period of the barycenter; thus, when phase-folding across multiple orbits, the occultations of the secondary planet will be smeared out \citep{kipping21a}, resulting in a transit light curve that is generally difficult to distinguish from an individual planet fit (Fig.~\ref{fig:transit}, bottom panels).

One possible binary planet detection approach that has not yet been considered extensively in the literature is measuring the Transit Duration Variation (TDV) signal (for a recent review, see \cite{agol18}). There are many physical mechanisms which can give rise to TDVs in exoplanetary systems, including orbital plane reorientation from multi-planet interactions or secular precession \citep{almenara15,mills17,boley20}, torques due to rotational oblateness of the host star \citep{szabo12,barnes13}, or eccentricity variations due to mean-motion resonance interactions \citep{nesvorny13}, making TDVs a powerful approach in the exoplanet observer's toolbox. Most relevant to our case of binary planets, however, is another known source of TDVs: the presence of a low-mass companion orbiting an exoplanet, i.e. an exomoon \citep{kipping08}. The larger satellite mass in our binary planet case results in a qualitative difference with the case of exomoons: whereas for an exomoon signal, the transit depth induced by the satellite is often not detectable, and thus the TTV/TDV is only a result of the slight gravitational tug of the satellite on the planet, for binary planets \textit{both} the planet and the satellite result in measurable transits. Yet more, their anticipated close separation leads to a single chimera transit (Figure~\ref{fig:transit}) whose TDVs (and TTVs) defy predictions for either component individually. As a result, the TTV/TDV effect is much more pronounced. The significant change in binary orbital phase from transit ingress to egress, and thus the varying radial extent of the binary, will vary greatly in each separate transit, giving rise to a TDV signal on the scale of hours---comparatively larger than most other known sources of TDVs (though not unheard of, e.g. circumbinary planets as in \cite{orosz12} or orbital plane reorientation near a grazing geometry as in \cite{hamann19}). And as with the case of exomoons, the TDV signal is also accompanied by a measurable Transit Timing Variation (TTV), which we model in this paper as another tool with which to distinguish binary planet systems.

In Section~\ref{sec:analytic}, we present an analytic equation for the transit duration of binary planet systems with arbitrary radii, separation, and period under the key assumptions of (i) 2-d motion, (ii) circular binary and stellarcentric orbits (relaxed slightly in Sec.~\ref{sec:analytic}.1), and (iii) binary orbit coplanar with the stellarcentric orbit; these assumptions, along with the derivation, are justified in Appendix A. We also provide a \python\ code for numerical computation of the transit duration and peak-to-peak TDV amplitudes, and compare our obtained results to transit simulations. Then, in Section~\ref{sec:timedomain} we present time-domain simulations of binary planets and analyze the TTV signals, examining the whole range of transit signatures left by these remarkable systems. Finally, we perform some rough calculations of the transit SNR required for detection of a binary planet via TDV, and point to a list of promising \textit{Kepler} candidates for further inspection.

\section{Analytic approximation for the transit duration} \label{sec:analytic}
\begin{table*}
\caption{Index of important variables used multiple times throughout the paper.}
\centering
\begin{tabular}{c c c}
\hline\hline
Symbol & Units\ & Definition \\ [0.5ex]
\hline
$a$ & distance & Stellarcentric orbital radius of the binary \\
$a_p$ & distance & Distance from binary barycenter to planet (physical units) \\
$\widetilde{a}_p$ & dimensionless (factors of $R_*$) & Distance from binary barycenter to planet (dimensionless units) \\
$\widetilde{a}$ & dimensionless (factors of $R_*$) & Stellarcentric orbital radius of the binary \\
$a_s$ & distance & Distance from binary barycenter to satellite (physical units) \\
$\widetilde{a}_s$ & dimensionless (units of $R_*$) & Distance from binary barycenter to satellite (dimensionless units) \\
$a_{sp}$ & dimensionless (factors of $p+s$) & Distance from planet to satellite, in factors of summed planet radii \\
$b$ & dimensionless & Impact parameter \\
$M_p$ & mass & Mass of planet \\
$M_s$ & mass & Mass of satellite \\
$p$ & dimensionless (factors of $R_*$) & Radius of planet \\
$P$ & time & Stellarcentric orbital period of binary barycenter \\
$P_s$ & time & Orbital period of planet and satellite about their barycenter \\
$R_{\mathrm{bin}}(t)$ & dimensionless (factors of $R_*$) & Instantaneous 2-d radius subtended by the binary at a given time \\
$R_*$ & distance & Stellar radius \\
$s$ & dimensionless (factors of $R_*$) & Radius of satellite \\ 
$T_{14}'$ & time & Transit duration \\
$t_e$ & time & Time of transit egress \\
$t_i$ & time & Time of transit ingress \\
$v_{\mathrm{bary}} $ & time$^{-1}$ & Speed of the barycenter; calculated with dimensionless distance \\
$\lambda_{sp}$ & dimensionless (angle) & Binary orbital phase (runs from 0 to $2\pi$) \\
$\omega$ & dimensionless (angle) & Argument of pericenter \\ [1ex]
\hline
\end{tabular}
\label{table:nonlin}
\end{table*}

\subsection{Case-by-case transit duration approximations} \label{subsec21}
As seen in Figure~\ref{fig:transit}, the overlapping transit signal from binary planets results in a light curve which can be difficult to distinguish from a conventional planetary transit, particularly in the case of infrequent sampling. Thus, in this work, we are interested in calculating the variations of the transit duration as a function of the phase angle $\lambda_{sp}$, which can then be extended to time domain simulations (Section~\ref{sec:timedomain}) to establish identifiable patterns in actual data. The transit duration waveform varies from a constant value as a result of the rapid rotation of the binary planets about their barycenter over the course of a transit. \cite{podsiadlowski10} and \cite{ochiai14} both found that these systems are stable only when their orbital radii are a small multiple ($\sim 2-5$x) of their summed radii, and thus their angular velocities about the barycenter will be large compared to the velocity of the barycenter about the star. Several transit waveforms are overplotted in Fig.~\ref{fig:transit} for the range of values $\lambda_{sp} \in [0, 2\pi]$ at ingress.

\begin{figure}
    \centering
    \includegraphics[width=0.75\linewidth]{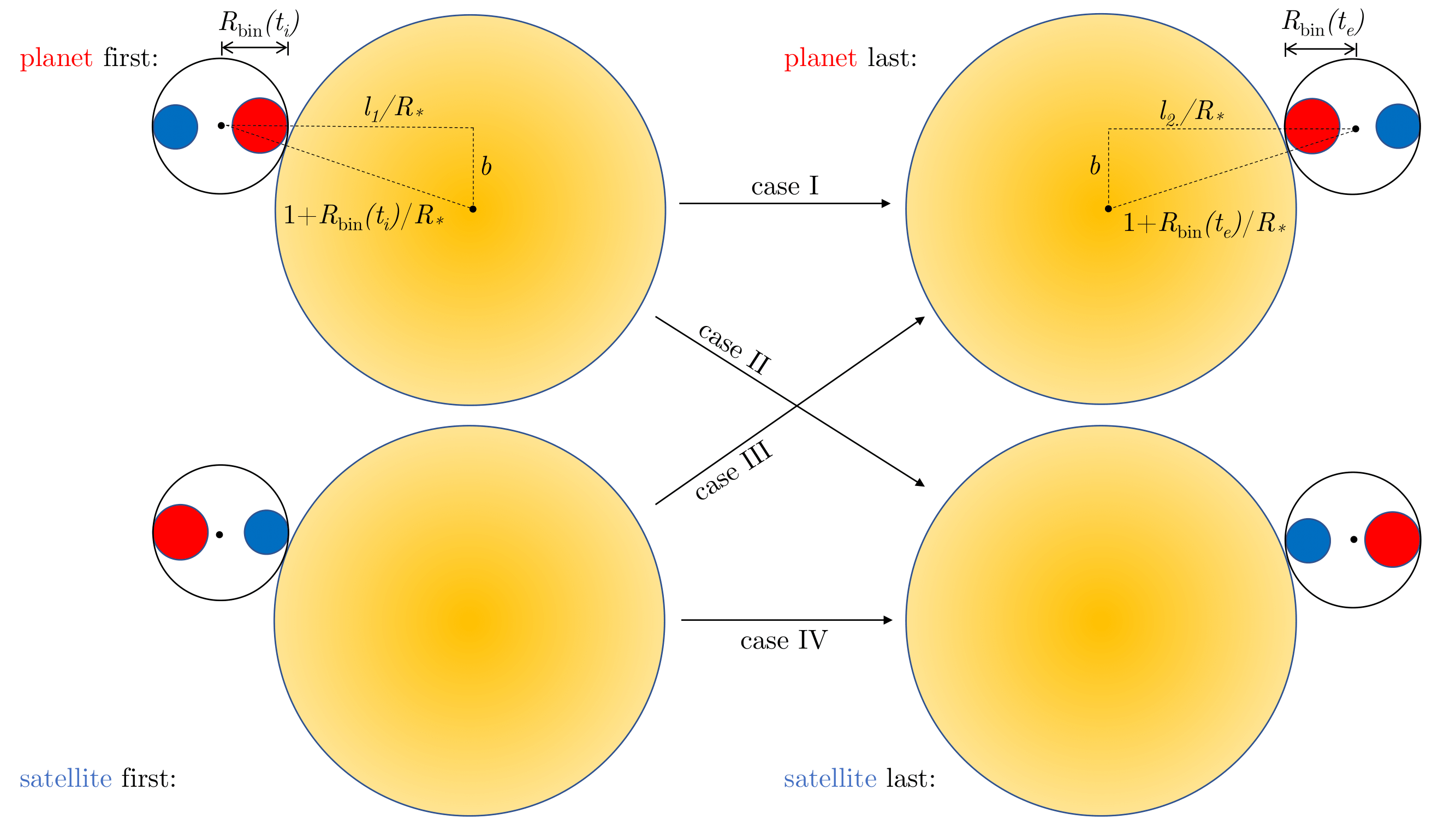}
    \caption{Possible configurations at $t_i$ (left) and $t_e$ (right). Red is the ``planet", blue is the ``satellite". Four possible transit cases arise from this construction, resulting in the piecewise form of Eqs.~\ref{eq:1} and ~\ref{eq:2}.}
    \label{fig:cases}
\end{figure}

For notational convenience, we define the more massive object as the planet, ``P", and the less massive object the satellite ``S" (this choice is irrelevant in the case of an exact binary). $\lambda_{sp}=0$ then corresponds to exact alignment of the planet and satellite as viewed by the observer, with the planet in front such that the satellite is completely obscured (i.e. inferior conjunction). $\lambda_{sp}=\pi$ corresponds to the inverse case with the satellite exactly in front of the planet, and $\lambda_{sp}=\pi/2$ and $\lambda_{sp}=3\pi/2$ correspond to cases of maximum orbital separation. The transit duration then becomes a piecewise function, with four separate cases (illustrated in Fig.~\ref{fig:cases}; these collapse into two cases for exact binaries):
\begin{enumerate}[I]
    \item Planet enters transit first, planet exits transit last.
    \item Planet enters transit first, satellite exits transit last.
    \item Satellite enters transit first, planet exits transit last.
    \item Satellite enters transit first, satellite exits transit last.
\end{enumerate}

Which of these cases is physically realized during a transit will be a function of the particular phase at ingress $\lambda_{sp}$, the angular orbital velocity of the binary planets about their barycenter, the separation radii ($a_s$, $a_p$), and the chimera transit's 1st-to-4th contact duration ($T_{14}'$)\footnote{Where we use the notation $T_{14}'$ rather than simply $T_{14}$ to indicate that this is the duration of the chimera transit, not the resolved transit duration of the planet or satellite individually.}. Because of the dependence on the latter parameter, $T_{14}'$, the piecewise equation for the transit duration will be transcendental, thus making an exact closed-form solution intractable. Still, in the limiting case of a equatorial transit ($b=0$) (which we relax later), we find that this transcendental equation has the form:

\begin{equation} \label{eq:1}
T_{14}'=
\begin{cases}
    \frac{P}{2\pi \widetilde{a}}\bigg[2 + 2p + \widetilde{a}_p\sin(\lambda_{sp}) - \widetilde{a}_p\sin\Big(\lambda_{sp}+\frac{2\pi T_{14}'}{P_s}\Big)\bigg] & \mathrm{case\;I} \\
    
    \frac{P}{2\pi \widetilde{a}}\bigg[2 + p + s + \widetilde{a}_p\sin(\lambda_{sp}) + \widetilde{a}_s\sin\Big(\lambda_{sp}+\frac{2\pi T_{14}'}{P_s}\Big)\bigg] & \mathrm{case\;II} \\
    
    \frac{P}{2\pi \widetilde{a}}\bigg[2 + p + s - \widetilde{a}_s\sin(\lambda_{sp}) - \widetilde{a}_p\sin\Big(\lambda_{sp}+\frac{2\pi T_{14}'}{P_s}\Big)\bigg] & \mathrm{case\;III} \\
    
    \frac{P}{2\pi \widetilde{a}}\bigg[2 + 2s - \widetilde{a}_s\sin(\lambda_{sp}) + \widetilde{a}_s\sin\Big(\lambda_{sp}+\frac{2\pi T_{14}'}{P_s}\Big)\bigg] & \mathrm{case\;IV} \\
\end{cases}
\end{equation}
where $P$ is the period of the binary planets' barycenter about the star, $\widetilde{a}$ is the corresponding orbital radius, $p$ and $s$ and the radii of the planet and satellite respectively, $\widetilde{a}_p$ and $\widetilde{a}_s$ are their orbital radii about the barycenter (all distances given in units of $R_*$, which is denoted by the tilde symbol), and $P_s$ (equivalently $P_p$) is the period of the binary planets about the barycenter, which is treated as following Kepler's Third Law as:

\begin{equation*}
P_s = 2\pi \bigg( \frac{(a_s+a_p)^3}{G(M_p + M_s)} \bigg)^{1/2}
\end{equation*}

Eq.~\ref{eq:1} is derived explicitly in Appendix A, but we include a broad-strokes description of the procedure here. Working entirely in a 2-d, face-on view of the transit, and assuming $b=0$ for now, we model the two-planet system as being contained ``within" one larger sphere of a variable radius equal to the total radial extent of the binary (as in Fig.~\ref{fig:cases}). For $b=0$, the larger sphere enters and exits transit exactly at the same time as the binary. Then, by writing out the expressions for the radius of the larger sphere at ingress and egress, we can write a piecewise expression for $T_{14}'$ in much the same way as the standard derivation for transit duration, but using a variable planet radius. Note that we also assume zero eccentricity for the planet-satellite system itself, as tides will rapidly circularize the binary for such small separations \citep{ochiai14}.

The approximation of Equation~\ref{eq:1} will always overestimate $T_{14}'$ for non-zero impact parameters. For $b>0$, a better approximation is:
\begin{equation} \label{eq:2}
T_{14}'=
\begin{cases}
    \frac{P}{2\pi \widetilde{a}}\left[ 2\sqrt{(1+p)^2-b^2} + \widetilde{a}_p\sin(\lambda_{sp}) - \widetilde{a}_p\sin\left(\lambda_{sp}+\frac{2\pi T_{14}'}{P_s}\right) \right]  & \mathrm{case\;I} \\
    
    \frac{P}{2\pi \widetilde{a}}\left[\sqrt{(1+p)^2-b^2} + \sqrt{(1+s)^2-b^2} + \widetilde{a}_p\sin(\lambda_{sp}) + \widetilde{a}_s\sin\left(\lambda_{sp}+\frac{2\pi T_{14}'}{P_s}\right) \right]  & \mathrm{case\;II} \\
     
    \frac{P}{2\pi \widetilde{a}}\left[\sqrt{(1+s)^2-b^2} + \sqrt{(1+p)^2-b^2} - \widetilde{a}_s\sin(\lambda_{sp}) - \widetilde{a}_p\sin\left(\lambda_{sp}+\frac{2\pi T_{14}'}{P_s}\right) \right]  & \mathrm{case\;III} \\
    
    \frac{P}{2\pi \widetilde{a}}\left[ 2\sqrt{(1+s)^2-b^2} - \widetilde{a}_s\sin(\lambda_{sp}) + \widetilde{a}_s\sin\left(\lambda_{sp}+\frac{2\pi T_{14}'}{P_s}\right) \right]  & \mathrm{case\;IV} \\
\end{cases}
\end{equation}

\begin{figure}
    \centering
    \includegraphics[width=\linewidth,keepaspectratio]{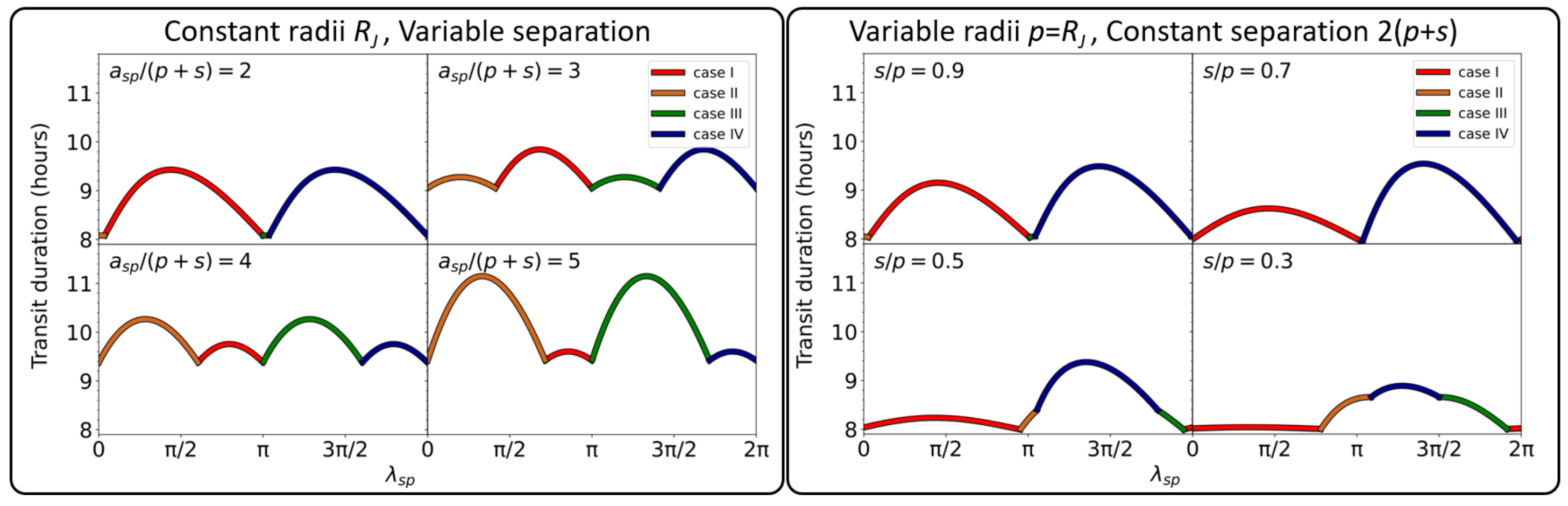}
    \caption{Variable transit durations of binary Jupiters around a Sun-like star, calculated from Eq.~\ref{eq:2} using the numerical solver provided at \href{https://github.com/joheenc/binary-planet-transits}{github.com/joheenc/binary-planet-transits}. \textbf{Left:} cases of exact binary Jupiters. The separation (in units of the summed planet radii) is varied from 2 to 5. The symmetry arising from equal radii and masses of the planet and satellite is manifestly visible in these plots. \textbf{Right:} cases with variable radius ratios $s/p$ (with $p$ fixed at $R_J$, $M_J$) and separation fixed to $2(p+s)$. The planet mass is scaled down assuming constant desnity. Notably, the TDV effects occur on the scale of $\sim$hours, much larger than most other known source of TDVs, with some exception e.g. circumbinary planets or orbital plane reorientation in grazing geometries.}
    \label{fig:tdur}
\end{figure}
For a full derivation of the impact parameter corrections, see Appendix B. The correction terms are derived by approximately subtracting off the differences in position between the larger sphere and the planet/satellite themselves. We note, as a check of the analytical result, that upon taking the limiting case of a single planet in Eq.~\ref{eq:2}---i.e., $\widetilde{a}_p=\widetilde{a}_s=0$ and $s=p$---we exactly recover the usual solution for circular orbits, Eq. (16) from \cite{seager03}.

Furthermore, in general the stellarcentric orbit of the barycenter will not be perfectly circular. A simple, first-order correction for this can be taken from \cite{winn10}: for nonzero eccentricity $e$ and argument of pericenter $\omega$, first replace the impact parameter $b$ with the eccentricity-corrected version, then multiply Eq.~\ref{eq:1} and \ref{eq:2} by a dimensionless factor:

\begin{equation*}
b \rightarrow b\bigg(\frac{1-e^2}{1+e\sin\omega}\bigg) \;\;\;\;\; \mathrm{and}\;\;\;\; T_{14}' \rightarrow T_{14}' \bigg(\frac{\sqrt{1-e^2}}{1+e\sin\omega}\bigg)
\end{equation*}
Both changes need to be made for systems with non-negligible stellarcentric eccentricity, i.e. recalculate $T'_{14}$ with the updated $b$ and then multiply by the above dimensionless factor. This factor accounts for the altered speed of the planet at conjunction, as in a non-circular orbit we may no longer assume a constant speed.

Both Equations~\ref{eq:1} and Eq.~\ref{eq:2} are transcendental for $T_{14}'$--a result of writing the radius of the larger sphere at egress as a function of $T_{14}'$ itself--and therefore cannot be accurately approximated by simple analytical methods. This stems from the fact the final phase position of the binary, which affects the duration, will itself depend on how many revolutions the binary performed, which in turn depends on the duration. Accordingly, we now turn to numerical solvers to obtain values for $T_{14}'$ as a function of the binary system parameters. We have provided a simple \python\ library for calculations of $T_{14}'$, available at \href{https://github.com/joheenc/binary-planet-transits}{github.com/joheenc/binary-planet-transits}. Fig.~\ref{fig:tdur} plots some sample $T_{14}'$ waveforms calculated numerically with this code.

We also note that these equations yield a relationship between the transit duration and the phase of the binary planets at ingress. However, because one does not know \textit{a-priori} the entry phase for each transit, reconstructing a full $T_{14}'$ waveform from several orbits of real transit data is not possible. Thus we turn to measurements of the peak-to-peak transit duration variation amplitude, i.e. the TDV signal, as computed from the total waveform. 
For a sample system of binary Jupiters at a orbital radii of 0.3-1 AU, these TDVs occur on the scale of hours.

In the large-separation, high-inclination limit, it is possible for the planet and satellite to transit separately in the light curve, complicating detection via TTV/TDV for single chimera transits. However, we show that this does not occur in cases of interest considered in this paper. The condition representing transit separation is:
\begin{equation*}
2\sqrt{1-b^2} \lesssim \widetilde{a}_p + \widetilde{a}_s + p + s
\end{equation*}
Mathematically, this represents the planet (satellite) leaving transit just as the satellite (planet) enters for the extreme case of maximum separation. For $b=0.5$, the highest impact parameter we consider in this paper (as this is the limit for the validity of the \cite{winn10} correction), we require $\widetilde{a}_p+\widetilde{a}_s+p+s \gtrsim 1.73$. In the case of exact binary Jupiters, for example, this would require $0.2 \times (1+a_{sp}) \gtrsim 1.73$, or equivalently $a_{sp}=7.65$. Even though we have taken extreme cases of both phase-dependent separation and detectable impact parameter, separation of planet and satellite transits would require a separation higher than the maximum value found in simulations of stable binary Jupiters formed via tidal interaction ($a_{sp} = 5$, \cite{ochiai14}). Geometrically, we expect such binaries to always have overlapping transits then.

In sufficiently high quality data, the chimera nature of the transit may still be distinguishable despite their overlap (e.g. the exomoon candidate Kepler-1708b-i displays such behaviour; \citealt{kipping22}). However, in automated searches, the duration of this combined chimera transit signal is easily measured and cataloged. In this way, we envisage the large chimera TDVs as being a simple way to quickly scour a large number of signals for potential candidates warranting more detailed photometric modeling. Moreover, because we have assumed binary planet orbits coplanar with the stellarcentric orbits, in some transit configurations (cases I and IV described above) there will be a mutual event in which one planet occults the other, causing a \textit{rise} in flux during the transit. Our transit duration derivation still holds for such brightening events, and these could be used as an additional signature to determine a deviation from a single-planet fit. In fact, such behavior has already been considered for mutual events of multi-planet systems \citep{ragozzine10} and observed to impart distinct signatures on the light curves \citep{masuda14,luger17} and spectra \citep{hirano12}, so in principle should be detectable for our binary planet case as well.

\subsection{Comparison of numerical TDV approximations to \luna\ dynamical simulations}

We compared our numerical approximations of the peak-to-peak amplitude calculated from our analytic approximation against dynamical modelling of binary planet transits using the \luna\ algorithm \citep{kipping11a}. \luna\ is an efficient analytic code initially designed for generating dynamic planet-moon transits, which is easily extended to our case of planet-planet transits. The algorithm accounts for limb darkening and all orbital elements (eccentricity, longitude of the ascending node, etc.), modelling the reflex motion of the planet due to the satellite at each time step.

\begin{figure}
    \centering
    \includegraphics[width=\linewidth,keepaspectratio]{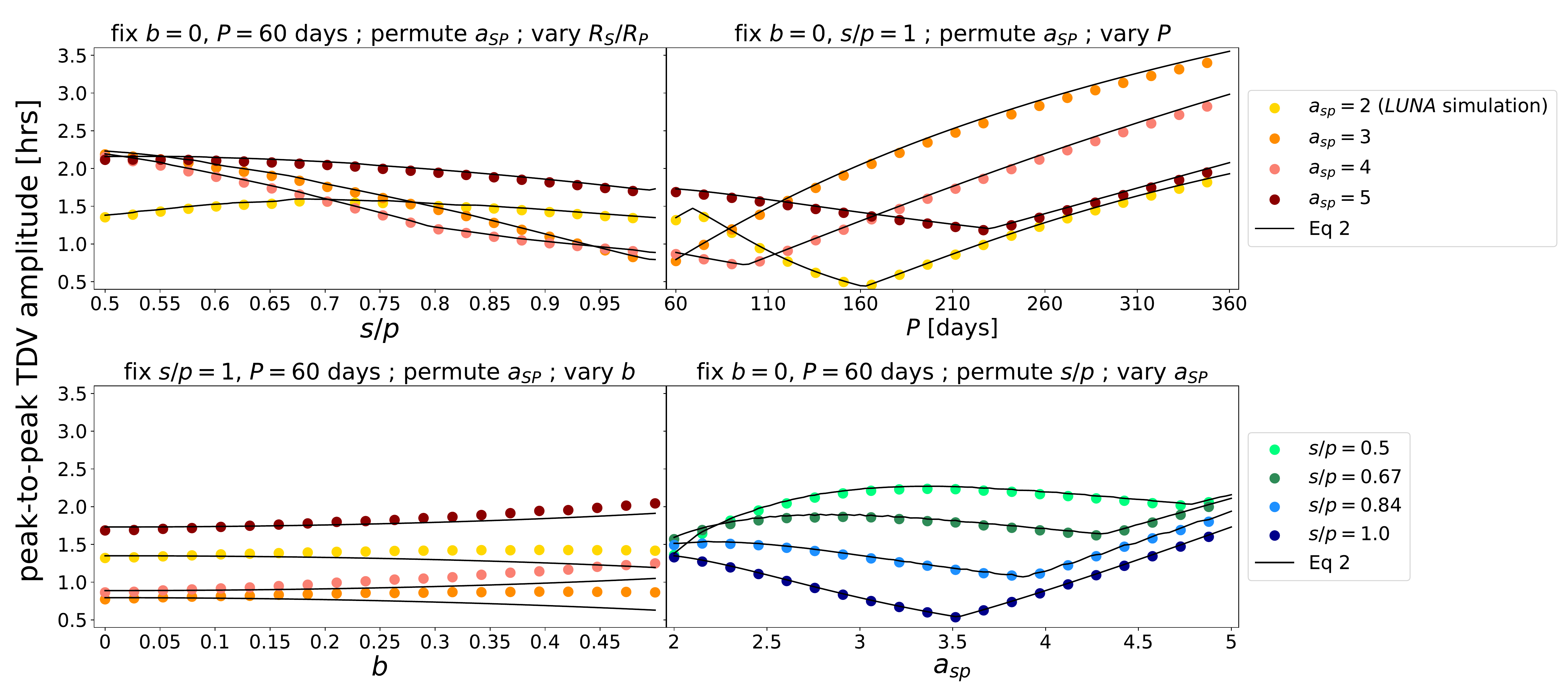}
    \caption{Verification of the analytic approximation in Eq.~\ref{eq:2} (black lines) against dynamical transit simulations using \luna\ (colored dots) across a wide range of physical parameters. In each panel we vary some parameter for four trial values of $a_{sp}$ (panels 1, 2, and 3) or $s/p$ (panel 4). The agreement is almost exact for nearly all explored regions of parameter space, with slight deviations likely occurring due to limb-darkening effects or discrete time steps. The exception is of higher values of $b$, at which our approximation breaks down. For all planets we assume Jupiter density.}
    \label{fig:luna}
\end{figure}

The agreement of \luna\ results with our analytic formulation to within 10\% within the physical parameter space for binary gas giants provides an independent confirmation of the validity of our results. At low $b$, the small deviations likely occur due to limb-darkening effects or discrete time steps. At higher $b$, our approximation of treating the two planets as one larger body breaks down, and so the larger deviation is expected. Moreover, since we have not made any limiting assumptions about the planet radii or separation in developing our approximation, in principle one could apply the same equations to, e.g. binary terrestrial planets (though these would likely result in smaller-amplitude TDV effects, and formation mechanisms for these systems are unexplored). Figure~\ref{fig:luna} overplots the \luna\ simulation results with our numerically computed peak-to-peak TDV amplitudes using the code provided in Section \ref{subsec21}.

\section{Time-domain simulations and Detectability with \textit{Kepler}}
\label{sec:timedomain}
In addition to large-amplitude TDV signals, we find that binary gas giants are also associated with a significant TTV effect, typically of order $\sim$tens of minutes. As with our TDV discussion, it should be emphasized that these TTVs are not the TTVs of any individual component in the system, but of the chimera signal itself. We calculate the flux weighted transit mid-time of our simulated chimera transits using

\begin{align}
t_{\mathrm{mid}} &= \frac{ \sum_{i=1}^N (1-f_i)t_i }{ \sum_{i=1}^N (1-f_i) }.
\end{align}

\begin{figure}
    \centering
    \includegraphics[width=\linewidth,keepaspectratio]{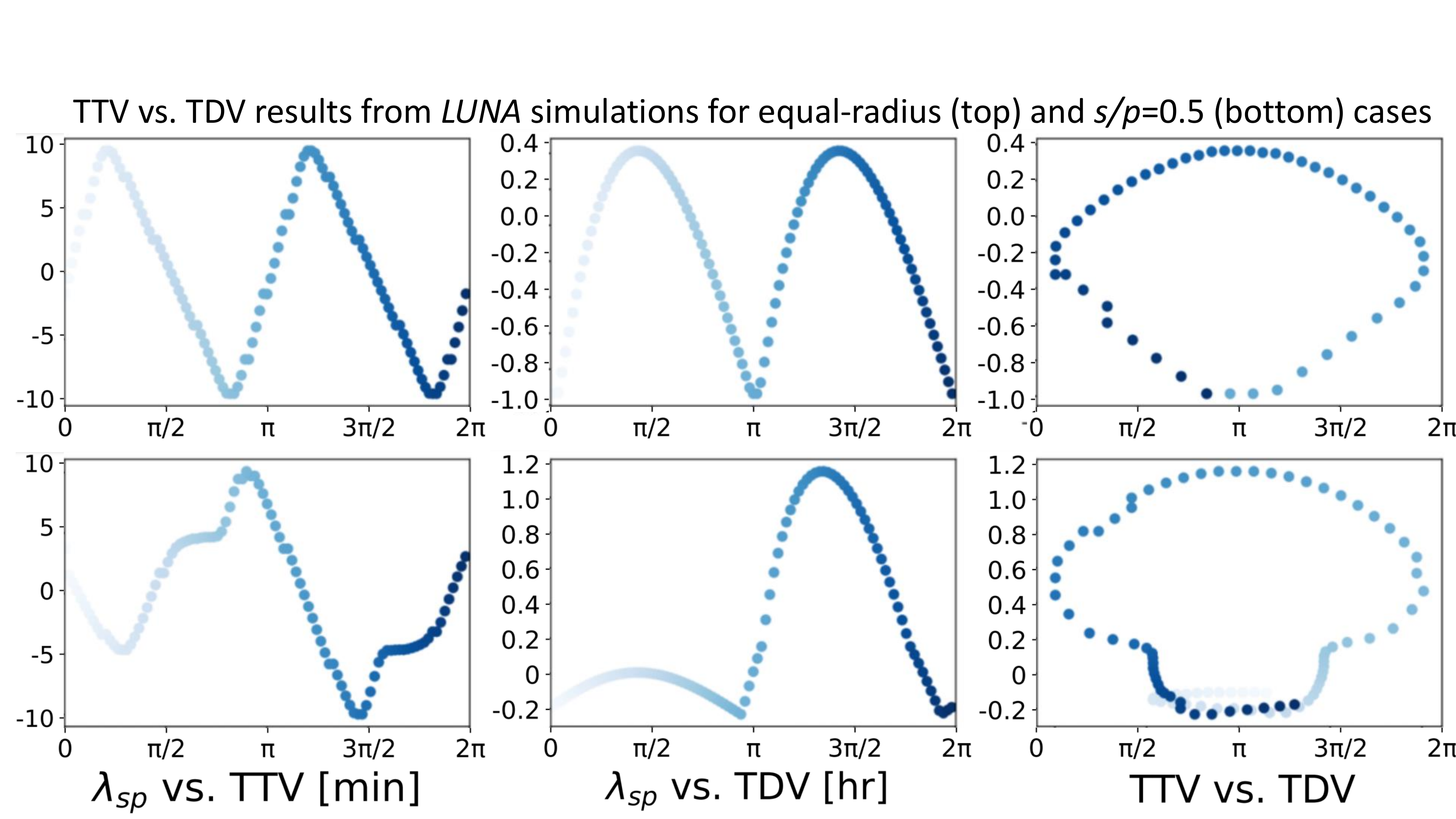}
    \caption{TTV and TDV amplitudes as functions of $\lambda_{sp}$ obtained from \textit{LUNA} dynamical simulations, alongside TTV-TDV curves. In the symmetric (top) case of binary Jupiters separated by twice the summed radii, we see the familiar elliptical path \citep{heller16} traced out in TTV-TDV space. However, this symmetry is broken for general, asymmetric binary planet systems, as seen in the bottom panel with $s/p=0.5$ and $a_{sp}=3$.}
    \label{fig:ttv}
\end{figure}

Figure~\ref{fig:ttv} plots the TTV and TDV amplitudes over $\lambda_{sp}\in[0, 2\pi)$ for sample systems of symmetric binary Jupiters and asymmetric binary planets with a 0.5 radius-ratio. While the standard elliptical TTV-TDV curve \citep{heller16} is roughly traced out in the symmetric case, in general we do not expect such symmetric behavior, as shown in the latter case. We simulated several sample systems to develop an intuition for the range of expected TTV-TDV behavior across the physical parameter space; the results are discussed in Appendix E and Fig.~\ref{fig:ttv-tdv_gallery}.

\begin{figure}
    \centering
    \includegraphics[width=\linewidth,keepaspectratio]{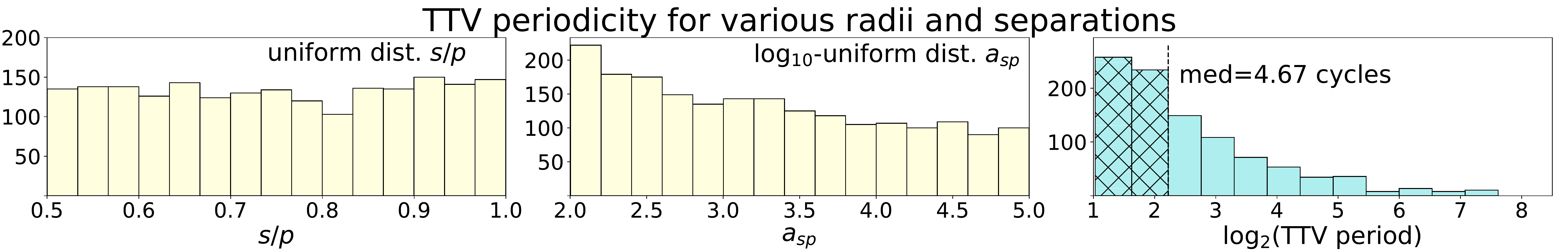}
    \caption{TTV periodicity measured by applying the Fast Fourier Transform. 2000 systems are drawn from a uniform distribution in $s/p \in [0.5, 1]$ (assuming Jupiter density for all planets), and a log-uniform distribution in $a_{sp}\in[2, 5]$, resulting in a similar distribution of TTV periodicities with half of measured periods falling between $\sim 2-4.7$ cycles. This hatched region represents the ``exomoon corridor" \citep{kipping21b}.}
    \label{fig:exomoon_corridor}
\end{figure}

To determine the TTV periodicity, we drew 2000 sample systems from a uniform distribution in $R_S/R_P$ between 0.5 and 1, and a log-uniform distribution in $a_{SP}$ between 2 and 5 times the summed planetary radii (motivated by findings from \cite{ochiai14}). We then ran 200 orbits of each sample system with \luna, and used the Fast Fourier Transform (FFT) algorithm to measure the TTV periodicity in planetary cycles. We find the TTV signal is of high measured frequency, falling within the recently identified exomoon corridor \citep{kipping21b}. In the case of exomoons, nearly half of all measured TTVs are expected to fall within a narrow period range of 2-4 cycles; for our sample runs, we find half of TTVs between 2 and 4.67 cycles (Figure~\ref{fig:exomoon_corridor}), in close agreement with the exomoon result (see also \citealt{teachey21}).

\subsection{TTV signal associated with Binary Planets}

The result may at first be surprising since the TTVs considered in \cite{kipping21b} are from small exomoons and ultimately driven by a different underlying mechanism: with exomoons, TTVs result from the gravitational tug of the moon on its host planet, but the flux dip from the moon itself is generally too small to result in a significant shift of the transit mid-time as calculated by the flux-weighted average. In the case of binary planets, both the primary planet and its satellite cause observable flux dips, resulting in a more dramatic shift of the measured transit mid-times between successive cycles. Nevertheless, the binary planet signal falls within the exomoon corridor for the same statistical reason as with exomoons: the TTV is of high frequency due to the rapid orbital period of the binary (on the scale of days) compared to the stellarcentric orbital period (on the scale of $\sim$dozens to hundreds of days). Thus, the TTV signal will always be undersampled due to having a frequency higher than the Nyquist rate, and the signal becomes aliased. The exomoon corridor can then be thought to apply generically to any high-frequency, undersampled source of TTVs; however, these are rare among currently known astrophysical sources, for now only belonging to exomoons and binary planets.

In Appendix Fig.~\ref{fig:ttv-tdv_gallery} we include TTV, TDV, and joint plots for 9 trial systems in the range of $R_S/R_P=0.5$-1 and $a_{SP}=2$-4 for transits simulated using \textit{LUNA}. We note some unusual properties of the TTV-TDV plots (third column), namely (I) they diverge significantly from the traditional elliptical shape of e.g. exomoon TTV-TDV curves \citep{heller16} and (II) while symmetry about the TTV axis is maintained, there is a symmetry-breaking for the TDV axis. For the case of TTVs, symmetry is expected, as we are calculating flux-weighted transit midtimes, and thus each rise in flux during transit ingress is mirrored by a corresponding occultations during egress (the same is true for occultations which occur after/before ingress/egress). However, for \textit{duration} variations, the asymmetric behavior is manifestly visible in the TDV-phase (column 2) plots, with discontinuous derivatives at each intersection of piecewise cases. As we are calculating transit durations as time of first-to-fourth contact, for asymmetric binary planets the larger planet will preferentially eclipse the star for a larger fraction of entry phases. These unique, half-symmetric TTV-TDV curves make for another observational signature in searching for potential binary planet systems.

\subsection{Detectability with \textit{Kepler}}

\cite{holczer16} catalogs the timings of 2599 \textit{Kepler} Objects of Interest (KOIs), of which 779 are sufficiently high-SNR to derive transit timing, duration, and depth. Of these, 260 show significant TTVs with long-term variations. For a rough calculation of the SNR required for a robust detection of a binary planet-induced TDV, we begin from the empirical result of \cite{holczer16} (see their Figure 2), a remarkably coherent inverse linear relationship between transit SNR and transit timing uncertainty:
\begin{equation*}
\sigma_{\mathrm{tmid}} = (\mathrm{100\;min}) / \mathrm{SNR}
\end{equation*}

For a trapezoidal transit, \cite{carter08} used Fisher covariance analysis to derive the following analytic approximations for uncertainty on transit timing and duration as a function of SNR:
\begin{equation*}
\sigma_{\mathrm{tmid}} = \mathrm{SNR}^{-1}T'_{14}\sqrt{\frac{\tau}{2T}}
\end{equation*}
\begin{equation*}
\sigma_{T} = \mathrm{SNR}^{-1}T'_{14}\sqrt{\frac{2\tau}{T}}
\end{equation*}
with $T$ is the transit duration as measured from the planet's center entering to exiting the stellar disk and $\tau$ the average duration of ingress and egress. Although $T \neq T'_{14}$, the uncertainties are similar and thus we adopt this in what follows. Thus, combining the two equations, we expect the uncertainty on transit duration to scale as $\sigma_{T} = 2\sigma_{\mathrm{tmid}} \simeq (\mathrm{200\;min})/(\mathrm{SNR})$.

We now turn to detectability of a typical binary planet system using this approximation. For our example cases of binary Jupiter-like planets orbiting 0.3 AU from a Sun-like star, we see average transit durations of $\sim 10$ hours, with a root-mean-squared TDV amplitude of 0.943 hours (Figure~\ref{fig:tdur}). For a robust 10$\sigma$ detection of the TDV signal, measurements satisfying $\sigma_{T'_{14}} < 0.943\mathrm{\;hr}/10 = 5.6{\mathrm{\;min}}$ would suffice, thus yielding a required transit SNR$=36$.

We filtered the results of \cite{holczer16} for planets with SNR$>36$ and orbital period $>60$ days, yielding a total of 113 candidates which could be inspected on a case-by-case basis for evidence of binary planet-like TTV and TDV behavior (these are listed in \href{https://github.com/joheenc/binary-planet-transits}{github.com/joheenc/binary-planet-transits}). To our knowledge, there are no examples of hour-amplitude, short-period TDV signals at this time, but our simple estimate here demonstrates that they are in principle detectable with current facilities. A detailed effort to measure the upper limit on the occurrence rate of such binary planets is beyond the scope of this paper, but is eminently possible and a subject for future work.

\section{Conclusion} \label{sec:conclusion}
Our work argues that binary planets should be detectable via large-amplitude TDVs. To this end, we have investigated the following:
\begin{itemize}
    \item An analytic approximation of the transit duration of binary planets, under the assumptions of a stellarcentric orbit coplanar with the binary planet orbit. Our approximation for the transit duration is exact for impact parameter $b=0$ (Equation~\ref{eq:1}), and accurate up to $b\lesssim 0.5$ (Equation~\ref{eq:2}). In the case of binary planet orbits misaligned with the stellarcentric orbit, we expect the TDV amplitude to decrease by some inclination-dependent factor; thus our work presents an upper limit on the expected TDVs.
    \item A \python\ code for numerical computation of the transit duration as a function of orbital phase ($\lambda_{sp}$), as well as the peak-to-peak Transit Duration Variation amplitude, at \href{https://github.com/joheenc/binary-planet-transits}{github.com/joheenc/binary-planet-transits}
    \item Identification of 113 high-SNR KOIs (also in the above GitHub) from the \cite{holczer16} Kepler TTV catalog; these sources make for promising candidates for an initial inspection for the binary planet TTV/TDV behavior.
    \item Identification of a significant short-period TTV signal, falling in the exomoon corridor of \cite{kipping21b}, as well as TTV-TDV curves (Fig~\ref{fig:ttv}, Fig~\ref{fig:ttv-tdv_gallery}) for a grid of sample systems to contribute another observational signature of binary planets.
\end{itemize}

To verify that our approximation gives valid results, we have compared against simulated transit durations and TDV amplitudes using the \luna\ dynamical simulation code, and found that the results agree to within 10\% in the region of parameter space explored. Most notably, we have found that binary gas giant transits exhibit TDVs on the scale of hours---significantly larger than other known sources, making them a promising candidate for detection via TDV. Moreover, we find that the TDVs are also accompanied by significant TTVs, and that the high-frequency TTV signal falls within the exomoon corridor. Binary planets thus join exomoons in the 2-4 cycle TTV period range, much lower than other known sources of timing variation, and thus marking a strong indicator that a system may be host to an orbiting companion around the larger-mass planet.

\section*{Acknowledgements}
We thank the anonymous referee for many extremely helpful comments which improved the quality of the paper. Special thanks to donors to the Cool Worlds Lab: Mark Sloan, Methven Forbes, Douglas Daughaday, Andrew Jones, Elena West, Tristan Zajonc, Chuck Wolfred, Lasse Skov, Graeme Benson, Alex de Vaal, Mark Elliott, Stephen Lee, Zachary Danielson, Chad Souter, Marcus Gillette, Tina Jeffcoat, Jason Rockett, Scott Hannum, Tom Donkin, Andrew Schoen, Jacob Black, Ms. Reza Ramezankhani, Steven Marks, Gary Canterbury, Nicholas Gebben, Mike Hedlund, Dhruv Bansal, Jonathan Sturm, Rand Corporation, Leigh Deacon, Ryan Provost, Brynjolfur Sigurjonsson, Benjamin Paul Walford, and Nicholas De Haan.

\section*{Data Availability}
Code and data produced during the preparation of this manuscript is available at \href{https://github.com/joheenc/binary-planet-transits}{github.com/joheenc/binary-planet-transits}.

\bibliography{manuscript}{}

\begin{thebibliography}{}
\makeatletter
\relax
\def\mn@urlcharsother{\let\do\@makeother \do\$\do\&\do\#\do\^\do\_\do\%\do\~}
\def\mn@doi{\begingroup\mn@urlcharsother \@ifnextchar [ {\mn@doi@}
  {\mn@doi@[]}}
\def\mn@doi@[#1]#2{\def\@tempa{#1}\ifx\@tempa\@empty \href
  {http://dx.doi.org/#2} {doi:#2}\else \href {http://dx.doi.org/#2} {#1}\fi
  \endgroup}
\def\mn@eprint#1#2{\mn@eprint@#1:#2::\@nil}
\def\mn@eprint@arXiv#1{\href {http://arxiv.org/abs/#1} {{\tt arXiv:#1}}}
\def\mn@eprint@dblp#1{\href {http://dblp.uni-trier.de/rec/bibtex/#1.xml}
  {dblp:#1}}
\def\mn@eprint@#1:#2:#3:#4\@nil{\def\@tempa {#1}\def\@tempb {#2}\def\@tempc
  {#3}\ifx \@tempc \@empty \let \@tempc \@tempb \let \@tempb \@tempa \fi \ifx
  \@tempb \@empty \def\@tempb {arXiv}\fi \@ifundefined
  {mn@eprint@\@tempb}{\@tempb:\@tempc}{\expandafter \expandafter \csname
  mn@eprint@\@tempb\endcsname \expandafter{\@tempc}}}

\bibitem[\protect\citeauthoryear{{Agol} \& {Fabrycky}}{{Agol} \&
  {Fabrycky}}{2018}]{agol18}
{Agol} E.,  {Fabrycky} D.,  2018, \mn@doi [Handbook of Exoplanets. Springer,
  Cham.] {10.1007/978-3-319-55333-7_7}

\bibitem[\protect\citeauthoryear{{Agol}, {Jansen}, {Lacy}, {Robinson}  \&
  {Meadows}}{{Agol} et~al.}{2015}]{agol15}
{Agol} E.,  {Jansen} T.,  {Lacy} B.,  {Robinson} T.~D.,   {Meadows} V.,  2015,
  \mn@doi [\apj] {10.1088/0004-637X/812/1/5}, \href
  {https://ui.adsabs.harvard.edu/abs/2015ApJ...812....5A} {812, 5}

\bibitem[\protect\citeauthoryear{{Akeson} et~al.,}{{Akeson}
  et~al.}{2013}]{akeson13}
{Akeson} R.~L.,  et~al., 2013, \mn@doi [\pasp] {10.1086/672273}, \href
  {https://ui.adsabs.harvard.edu/abs/2013PASP..125..989A} {125, 989}

\bibitem[\protect\citeauthoryear{{Almenara}, {Díaz}  \& {Mardling}}{{Almenara}
  et~al.}{2015}]{almenara15}
{Almenara} J.,  {Díaz} R.,   {Mardling} R.~{\textit{et al}}.,  2015, \mn@doi
  [MNRAS] {10.1093/mnras/stv1735}, 453, 2644

\bibitem[\protect\citeauthoryear{{Barnes} \& {O'Brien}}{{Barnes} \&
  {O'Brien}}{2002}]{barnes02}
{Barnes} J.~W.,  {O'Brien} D.~P.,  2002, \mn@doi [\apj] {10.1086/341477}, \href
  {https://ui.adsabs.harvard.edu/abs/2002ApJ...575.1087B} {575, 1087}

\bibitem[\protect\citeauthoryear{{Barnes}, {van Eyken}  \& {Jackson}}{{Barnes}
  et~al.}{2013}]{barnes13}
{Barnes} J.,  {van Eyken} J.,   {Jackson} B.~{\textit{et al}}.,  2013, \mn@doi
  [ApJ] {10.1088/0004-637X/774/1/53}, 774, 53

\bibitem[\protect\citeauthoryear{{Boley}, {van Laerhoven}  \& {Granados
  Contreras}}{{Boley} et~al.}{2020}]{boley20}
{Boley} D.,  {van Laerhoven} C.,   {Granados Contreras} A.,  2020, \mn@doi [AJ]
  {10.3847/1538-3881/ab8067}, 159, 207

\bibitem[\protect\citeauthoryear{{Canup} \& {Ward}}{{Canup} \&
  {Ward}}{2006}]{canup06}
{Canup} R.~M.,  {Ward} W.~R.,  2006, \mn@doi [\nat] {10.1038/nature04860},
  \href {https://ui.adsabs.harvard.edu/abs/2006Natur.441..834C} {441, 834}

\bibitem[\protect\citeauthoryear{{Carter}, {Yee}  \& {Eastman}}{{Carter}
  et~al.}{2008}]{carter08}
{Carter} J.,  {Yee} J.,   {Eastman} J.~{\textit{et al}}.,  2008, \mn@doi [ApJ]
  {10.1086/592321}, 689, 499

\bibitem[\protect\citeauthoryear{{Chrenko}, {Bro{\v{z}}}  \&
  {Nesvorn{\'y}}}{{Chrenko} et~al.}{2018}]{chrenko18}
{Chrenko} O.,  {Bro{\v{z}}} M.,   {Nesvorn{\'y}} D.,  2018, \mn@doi [\apj]
  {10.3847/1538-4357/aaeb93}, \href
  {https://ui.adsabs.harvard.edu/abs/2018ApJ...868..145C} {868, 145}

\bibitem[\protect\citeauthoryear{{Delrez} et~al.,}{{Delrez}
  et~al.}{2021}]{delrez21}
{Delrez} L.,  et~al., 2021, \mn@doi [Nature Astronomy]
  {10.1038/s41550-021-01381-5}, \href
  {https://ui.adsabs.harvard.edu/abs/2021NatAs...5..775D} {5, 775}

\bibitem[\protect\citeauthoryear{{Domingos}, {Winter}  \&
  {Yokoyama}}{{Domingos} et~al.}{2006}]{domingos06}
{Domingos} R.~C.,  {Winter} O.~C.,   {Yokoyama} T.,  2006, \mn@doi [\mnras]
  {10.1111/j.1365-2966.2006.11104.x}, \href
  {https://ui.adsabs.harvard.edu/abs/2006MNRAS.373.1227D} {373, 1227}

\bibitem[\protect\citeauthoryear{{Ford} \& {Rasio}}{{Ford} \&
  {Rasio}}{2008}]{ford08}
{Ford} E.,  {Rasio} F.,  2008, \mn@doi [ApJ] {10.1086/590926}, 686, 621

\bibitem[\protect\citeauthoryear{{Hamann}, {Montet}, {Fabrycky}, {Agol}  \&
  {Kruse}}{{Hamann} et~al.}{2019}]{hamann19}
{Hamann} A.,  {Montet} B.~T.,  {Fabrycky} D.~C.,  {Agol} E.,   {Kruse} E.,
  2019, \mn@doi [\aj] {10.3847/1538-3881/ab32e3}, \href
  {https://ui.adsabs.harvard.edu/abs/2019AJ....158..133H} {158, 133}

\bibitem[\protect\citeauthoryear{{Hamers} \& {Portegies Zwart}}{{Hamers} \&
  {Portegies Zwart}}{2018}]{hamers18}
{Hamers} A.~S.,  {Portegies Zwart} S.~F.,  2018, \mn@doi [\apjl]
  {10.3847/2041-8213/aaf3a7}, \href
  {https://ui.adsabs.harvard.edu/abs/2018ApJ...869L..27H} {869, L27}

\bibitem[\protect\citeauthoryear{{Heller}, {Hippke}  \& {Placek}}{{Heller}
  et~al.}{2016}]{heller16}
{Heller} R.,  {Hippke} M.,   {Placek} B.~{\textit{et al}}.,  2016, \mn@doi
  [A\&A] {10.1051/0004-6361/201628573}, 591, id.A67

\bibitem[\protect\citeauthoryear{{Hirano} et~al.,}{{Hirano}
  et~al.}{2012}]{hirano12}
{Hirano} T.,  et~al., 2012, \mn@doi [\apjl] {10.1088/2041-8205/759/2/L36},
  \href {https://ui.adsabs.harvard.edu/abs/2012ApJ...759L..36H} {759, L36}

\bibitem[\protect\citeauthoryear{{Holczer}, {Mazeh}  \& {Nachmani}}{{Holczer}
  et~al.}{2016}]{holczer16}
{Holczer} T.,  {Mazeh} T.,   {Nachmani} G.~{\textit{et al}}.,  2016, \mn@doi
  [ApJS] {10.3847/0067-0049/225/1/9}, 225, 9

\bibitem[\protect\citeauthoryear{{Kipping}}{{Kipping}}{2008}]{kipping08}
{Kipping} D.,  2008, \mn@doi [MNRAS] {10.1111/j.1365-2966.2008.13999.x}, 392,
  181

\bibitem[\protect\citeauthoryear{{Kipping}}{{Kipping}}{2011}]{kipping11a}
{Kipping} D.,  2011, \mn@doi [MNRAS] {10.1111/j.1365-2966.2011.19086.x}, 416,
  689

\bibitem[\protect\citeauthoryear{{Kipping}}{{Kipping}}{2021a}]{kipping21b}
{Kipping} D.,  2021a, \mn@doi [MNRAS] {10.1093/mnras/staa3398}, 500, 1851

\bibitem[\protect\citeauthoryear{{Kipping}}{{Kipping}}{2021b}]{kipping21a}
{Kipping} D.,  2021b, \mn@doi [MNRAS] {10.1093/mnras/stab2013}, 507, 4120

\bibitem[\protect\citeauthoryear{{Kipping}}{{Kipping}}{2022}]{kipping22}
{Kipping} D.,  2022, \mn@doi [Nature Astronomy] {10.1038/s41550-021-01539-1},
  6, 367

\bibitem[\protect\citeauthoryear{{Lewis}, {Ochiai}  \& {Nagasawa}}{{Lewis}
  et~al.}{2015}]{lewis15}
{Lewis} K.,  {Ochiai} H.,   {Nagasawa} M.~{\textit{et al}}.,  2015, \mn@doi
  [ApJ] {10.1088/0004-637X/805/1/27}, 805, 27

\bibitem[\protect\citeauthoryear{{Luger} et~al.,}{{Luger}
  et~al.}{2017}]{luger17}
{Luger} R.,  et~al., 2017, \mn@doi [Nature Astronomy]
  {10.1038/s41550-017-0129}, \href
  {https://ui.adsabs.harvard.edu/abs/2017NatAs...1E.129L} {1, 0129}

\bibitem[\protect\citeauthoryear{{Masuda}}{{Masuda}}{2014}]{masuda14}
{Masuda} K.,  2014, \mn@doi [\apj] {10.1088/0004-637X/783/1/53}, \href
  {https://ui.adsabs.harvard.edu/abs/2014ApJ...783...53M} {783, 53}

\bibitem[\protect\citeauthoryear{{Mills}, {Fabrycky}  \& {Migaszewski}}{{Mills}
  et~al.}{2017}]{mills17}
{Mills} S.,  {Fabrycky} D.,   {Migaszewski} C.~{\textit{et al}}.,  2017,
  \mn@doi [AJ] {10.3847/1538-3881/153/1/45}, 153, 45

\bibitem[\protect\citeauthoryear{{Moraes} \& {Vieira Neto}}{{Moraes} \& {Vieira
  Neto}}{2020}]{moraes20}
{Moraes} R.~A.,  {Vieira Neto} E.,  2020, \mn@doi [\mnras]
  {10.1093/mnras/staa1441}, \href
  {https://ui.adsabs.harvard.edu/abs/2020MNRAS.495.3763M} {495, 3763}

\bibitem[\protect\citeauthoryear{{Nesvorný}, {Kipping}  \&
  {Terrell}}{{Nesvorný} et~al.}{2013}]{nesvorny13}
{Nesvorný} D.,  {Kipping} D.,   {Terrell} D.~{\textit{et al}}.,  2013, \mn@doi
  [ApJ] {10.1088/0004-637X/777/1/3}, 777, 3

\bibitem[\protect\citeauthoryear{{Ochiai}, {Nagasawa}  \& {Ida}}{{Ochiai}
  et~al.}{2014}]{ochiai14}
{Ochiai} H.,  {Nagasawa} M.,   {Ida} S.,  2014, \mn@doi [ApJ]
  {10.1088/0004-637X/790/2/92}, 790, 92

\bibitem[\protect\citeauthoryear{{Orosz}, {Welsh}  \& {Carter}}{{Orosz}
  et~al.}{2012}]{orosz12}
{Orosz} J.,  {Welsh} W.,   {Carter} J.~{\textit{et al}}.,  2012, \mn@doi
  [Science] {10.1126/science.1228380}, 337, 1511

\bibitem[\protect\citeauthoryear{{Podsiadlowski}, {Rappaport}  \&
  {Fregeau}}{{Podsiadlowski} et~al.}{2010}]{podsiadlowski10}
{Podsiadlowski} P.,  {Rappaport} S.,   {Fregeau} J.~{\textit{et al}}.,  2010,
  arXiv:1007.1418

\bibitem[\protect\citeauthoryear{{Quarles}, {Eggl}, {Rosario-Franco}  \&
  {Li}}{{Quarles} et~al.}{2021}]{quarles21}
{Quarles} B.,  {Eggl} S.,  {Rosario-Franco} M.,   {Li} G.,  2021, \mn@doi [\aj]
  {10.3847/1538-3881/ac042a}, \href
  {https://ui.adsabs.harvard.edu/abs/2021AJ....162...58Q} {162, 58}

\bibitem[\protect\citeauthoryear{{Ragozzine} \& {Holman}}{{Ragozzine} \&
  {Holman}}{2010}]{ragozzine10}
{Ragozzine} D.,  {Holman} M.~J.,  2010, arXiv e-prints, \href
  {https://ui.adsabs.harvard.edu/abs/2010arXiv1006.3727R} {p. arXiv:1006.3727}

\bibitem[\protect\citeauthoryear{{Rasio} \& {Ford}}{{Rasio} \&
  {Ford}}{1996}]{rasio96}
{Rasio} F.,  {Ford} E.,  1996, \mn@doi [Sci] {10.1126/science.274.5289.954},
  274, 954

\bibitem[\protect\citeauthoryear{{Rosario-Franco}, {Quarles}, {Musielak}  \&
  {Cuntz}}{{Rosario-Franco} et~al.}{2020}]{rosario20}
{Rosario-Franco} M.,  {Quarles} B.,  {Musielak} Z.~E.,   {Cuntz} M.,  2020,
  \mn@doi [\aj] {10.3847/1538-3881/ab89a7}, \href
  {https://ui.adsabs.harvard.edu/abs/2020AJ....159..260R} {159, 260}

\bibitem[\protect\citeauthoryear{{Sartoretti} \& {Schneider}}{{Sartoretti} \&
  {Schneider}}{1999}]{sartoretti99}
{Sartoretti} P.,  {Schneider} J.,  1999, \mn@doi [\aaps] {10.1051/aas:1999148},
  \href {https://ui.adsabs.harvard.edu/abs/1999A&AS..134..553S} {134, 553}

\bibitem[\protect\citeauthoryear{{Seager} \& {Mallén-Ornelas}}{{Seager} \&
  {Mallén-Ornelas}}{2003}]{seager03}
{Seager} S.,  {Mallén-Ornelas} G.,  2003, \mn@doi [ApJ] {10.1086/346105}, 585,
  1038

\bibitem[\protect\citeauthoryear{{Spalding}, {Batygin}  \& {Adams}}{{Spalding}
  et~al.}{2016}]{spalding16}
{Spalding} C.,  {Batygin} K.,   {Adams} F.~C.,  2016, \mn@doi [\apj]
  {10.3847/0004-637X/817/1/18}, \href
  {https://ui.adsabs.harvard.edu/abs/2016ApJ...817...18S} {817, 18}

\bibitem[\protect\citeauthoryear{{Stern} \& {Levison}}{{Stern} \&
  {Levison}}{2002}]{stern02}
{Stern} S.~A.,  {Levison} H.~F.,  2002, Highlights of Astronomy, \href
  {https://ui.adsabs.harvard.edu/abs/2002HiA....12..205S} {12, 205}

\bibitem[\protect\citeauthoryear{{Szabó}, {Pál}  \& {Derekas}}{{Szabó}
  et~al.}{2012}]{szabo12}
{Szabó} G.,  {Pál} A.,   {Derekas} A.~{\textit{et al}}.,  2012, \mn@doi
  [MNRAS: Letters] {10.1111/j.1745-3933.2012.01219.x}, 421, L122

\bibitem[\protect\citeauthoryear{{Teachey}}{{Teachey}}{2021}]{teachey21}
{Teachey} A.,  2021, \mn@doi [MNRAS] {10.1093/mnras/stab1840}, 506, 2104

\bibitem[\protect\citeauthoryear{{Vanderburg} \& {Rodriguez}}{{Vanderburg} \&
  {Rodriguez}}{2021}]{vanderburg21}
{Vanderburg} A.,  {Rodriguez} J.~E.,  2021, \mn@doi [\apjl]
  {10.3847/2041-8213/ac33b4}, \href
  {https://ui.adsabs.harvard.edu/abs/2021ApJ...922L...2V} {922, L2}

\bibitem[\protect\citeauthoryear{{Vanderburg}, {Rappaport}  \&
  {Mayo}}{{Vanderburg} et~al.}{2018}]{vanderburg18}
{Vanderburg} A.,  {Rappaport} S.~A.,   {Mayo} A.~W.,  2018, \mn@doi [\aj]
  {10.3847/1538-3881/aae0fc}, \href
  {https://ui.adsabs.harvard.edu/abs/2018AJ....156..184V} {156, 184}

\bibitem[\protect\citeauthoryear{{Winn}}{{Winn}}{2010}]{winn10}
{Winn} J.,  2010, \mn@doi [EXOPLANETS, University of Arizona Press]
  {10.48550/arXiv.1001.2010}

\makeatother
\end{thebibliography}
\bibliographystyle{mnras}

\appendix
\counterwithin{figure}{section}
\counterwithin{equation}{section}
\section{Derivation of transit duration}
We work with dimensionless distances given in units of $R_*$ (see Tab~\ref{table:nonlin} for an index of variables) and make the following assumptions:
\begin{itemize}
    \item 2-d motion over the course of a transit (i.e. constant barycenter velocity).
    \item The binary and stellarcentric orbits are both circular (we relax the latter assumption following the method adopted in \citealt{winn10}).
    \item The binary orbit is coplanar with the stellarcentric orbit. We adopt this assumption for the sake of simplicity, as writing down an expression for variable binary-stellarcentric obliquity is a more difficult problem, and the coplanar case considered in our paper is a first step towards a more complete treatement of binary planet TDVs. This is not to say that there cannot be binary planets with their orbital plane misoriented with respect to the stellarcentric orbit; we expect such systems to have their TDV amplitudes scaled down by some inclination-dependent factor, but the determination of this factor is beyond the scope of this work. Thus, the TDVs discussed in our paper (e.g. in Fig.~\ref{fig:luna}) are really upper limits on the TDV amplitude.
\end{itemize}

Let the moment at which the binary begins its transit (ingress) be $t_i$ and the moment at which the transit ends (egress) be $t_e$. There are four transit cases, defined in Sec.~\ref{sec:analytic}, corresponding to the planet or satellite leading ingress and the planet or satellite trailing egress. We can approximate the binary planets as one larger ``planet" with a center coincident with the binary barycenter and a radius $R_{\mathrm{bin}}(t)$ equal to the maximum radial extent of the two planets. This definition means the larger ``planet" has a variable radius, equal at $t_i$ to:
\begin{equation*}
R_{\mathrm{bin}}(t_i)=
\begin{cases}
    \widetilde{a}_p\sin(\lambda_{sp}) + p & \mathrm{cases\;I,II} \\
    -\widetilde{a}_s\sin(\lambda_{sp}) + s & \mathrm{cases\;III,IV}
\end{cases}
\end{equation*}
where $\lambda_{sp}$ is defined as the orbital phase of the binary at the moment of ingress (limiting cases of $\lambda_{sp}=0$, $\pi/2$, $\pi$, and $3\pi/2$ are defined in Sec.~\ref{sec:analytic}); and $\widetilde{a}_p$ and $\widetilde{a}_s$ are the separations of the planet (radius $p$) and satellite (radius $s$) from the binary barycenter.

Now, we can write $R_{\mathrm{bin}}(t_e)$ at egress in terms of the transit duration $T'_{14}$:
\begin{equation*}
R_{\mathrm{bin}}(t_e)=
\begin{cases}
    -\widetilde{a}_p\sin\left( \lambda_{sp} + 2\pi\frac{\mathrm{mod}(T'_{14}, P_s)}{P_s} \right) + p & \mathrm{cases\;I,III} \\
    \widetilde{a}_s\sin\left( \lambda_{sp} + 2\pi\frac{\mathrm{mod}(T'_{14}, P_s)}{P_s} \right) + s & \mathrm{cases\;II,IV}
\end{cases}
\end{equation*}
Note that in all cases within the physical parameter space, $T'_{14}$ is shorter than $P_s$, so $\mathrm{mod}(T'_{14}, P_s)$ will henceforth be written simply as $T'_{14}$.

We are now interested in calculating $l_1$ and $l_2$, the horizontal lengths spanned by the star and large ``planet" at times $t_i$ and $t_e$, respectively (Fig.~\ref{fig:cases}). These will be:
\begin{align*}
    l_1 = \sqrt{(1+R_{\mathrm{bin}}(t_i))^2-b^2} \\
    l_2 = \sqrt{(1+R_{\mathrm{bin}}(t_e))^2-b^2}
\end{align*}
Then we can write:
\begin{equation*}
\sin(\alpha_1) = \frac{l_1}{\widetilde{a}} \; ;\; \sin(\alpha_2) = \frac{l_2}{\widetilde{a}}
\end{equation*}
where $\alpha_1$ is the angle subtended by the barycenter of the binary from ingress to transit midpoint, $\alpha_2$ is the same angle subtended from transit midpoint to egress, and $\tilde{a}$ is the semimajor axis of the binary system around the star. Then we can write the transit duration:
\begin{equation*}
T'_{14} = P \frac{\alpha_1 + \alpha_2}{2\pi}
\end{equation*}
where $P$ is the orbital period of the binary barycenter around the star. Thus, we can write down a piecewise equation for $T'_{14}$: \\
\begin{equation} \label{eq:A1}
T'_{14}=
\begin{cases}
    \frac{P}{2\pi}\left[ \sin^{-1}\left(\frac{\sqrt{(1+\widetilde{a}_p\sin(\lambda_{sp})+p)^2-b^2}}{\tilde{a}} \right) + \sin^{-1}\left(\frac{\sqrt{\left(1-\widetilde{a}_p\sin\left(\lambda_{sp}+2\pi\frac{T'_{14}}{P_s}\right)+p\right)^2-b^2}}{\tilde{a}} \right) \right]-t_{\mathrm{cor,i}}-t_{\mathrm{cor,e}} & \mathrm{case\;I} \\
    \frac{P}{2\pi}\left[ \sin^{-1}\left(\frac{\sqrt{(1+\widetilde{a}_p\sin(\lambda_{sp})+p)^2-b^2}}{\tilde{a}} \right) + \sin^{-1}\left(\frac{\sqrt{\left(1+\widetilde{a}_s\sin\left(\lambda_{sp}+2\pi\frac{T'_{14}}{P_s}\right)+s\right)^2-b^2}}{\tilde{a}} \right) \right]-t_{\mathrm{cor,i}}-t_{\mathrm{cor,e}} & \mathrm{case\;II} \\
    \frac{P}{2\pi}\left[ \sin^{-1}\left(\frac{\sqrt{(1-\widetilde{a}_s\sin(\lambda_{sp})+s)^2-b^2}}{\tilde{a}} \right) + \sin^{-1}\left(\frac{\sqrt{\left(1-\widetilde{a}_p\sin\left(\lambda_{sp}+2\pi\frac{T'_{14}}{P_s}\right)+p\right)^2-b^2}}{\tilde{a}} \right) \right]-t_{\mathrm{cor,i}}-t_{\mathrm{cor,e}} & \mathrm{case\;III} \\
    \frac{P}{2\pi}\left[ \sin^{-1}\left(\frac{\sqrt{(1-\widetilde{a}_s\sin(\lambda_{sp})+s)^2-b^2}}{\tilde{a}} \right) + \sin^{-1}\left(\frac{\sqrt{\left(1+\widetilde{a}_s\sin\left(\lambda_{sp}+2\pi\frac{T'_{14}}{P_s}\right)+s\right)^2-b^2}}{\tilde{a}} \right) \right]-t_{\mathrm{cor,i}}-t_{\mathrm{cor,e}} & \mathrm{case\;IV} \\
\end{cases}
\end{equation}
where the $t_{\mathrm{cor,i}}$ and $t_{\mathrm{cor,e}}$ are small terms accounting for the overestimation effect of our large ``planet" approximation at nonzero impact parameters (derived in Appendix B below). Noting that $\alpha_1$, $\alpha_2$ will be small ($T'_{14} \ll P)$, we can Taylor approximate $\sin^{-1}(\frac{l_1}{\tilde{a}}) + \sin^{-1}(\frac{l_2}{\tilde{a}}) \approx \frac{l_1+l_2}{\tilde{a}}$ to first order and combine with the impact parameter correction terms derived in Appendix B to arrive at Eq.~\ref{eq:2}. Further assuming $b=0$ gives Eq.~\ref{eq:1}. Unfortunately, this is a transcendental equation for $T'_{14}$ which is difficult to approximate analytically. We thus defer to numerical codes for calculations involving these equations.

\section{Derivation of impact parameter correction terms}
Our approximation of the binary planets as one large ``planet" will overpredict transit durations for any nonzero impact parameter by erroneously assuming an earlier-than-physical time of first contact/ingress and later-than-physical time of last contact/egress. Here we derive a simple correction term to improve accuracy for small nonzero values of $b$.

\begin{figure}
    \centering
    \includegraphics[width=0.85\linewidth]{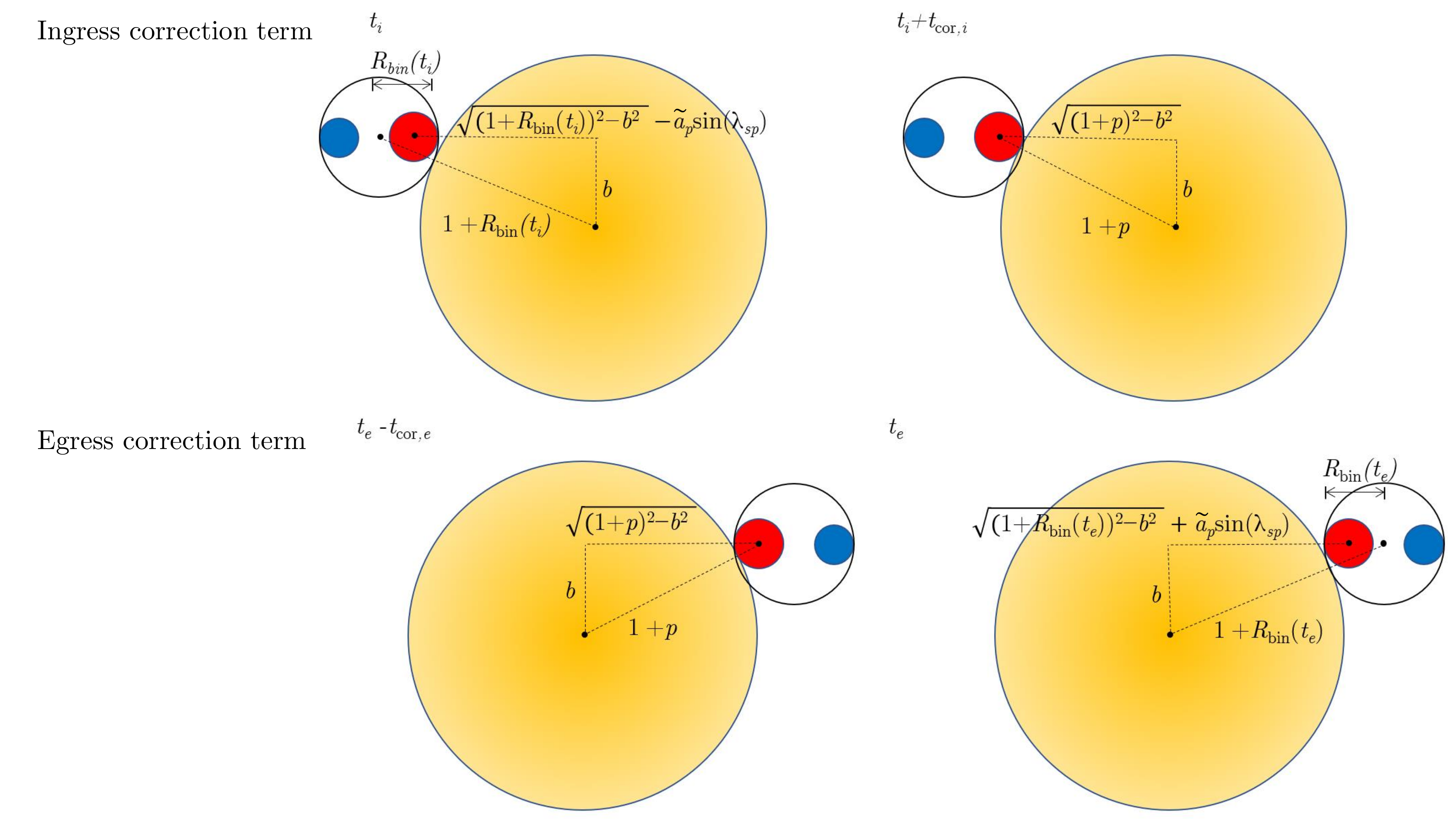}
    \caption{Guiding figure for the impact parameter correction terms derived in Appendix b. \textbf{Top:} the top-left panel shows the binary planet as the larger ``planet" is at ingress. Shortly after, the top-right panel shows the binary as the actual (red) planet is at ingress, i.e. the true first contact of the transit. The time between these two events is derived by assuming a constant orbital velocity and binary phase $\lambda_{sp}$. \textbf{Bottom:} a similar argument for egress.}
    \label{fig:impactcorr}
\end{figure}

First consider the transit ingress case. The leading planet travels some small distance during the elapsed time between $t_i$, the time of first contact of the large ``planet", and the actual beginning of the transit, the time of first contact of the leading body (the planet in cases I and II, the satellite in cases III and IV). Taking the phase $\lambda_{sp}$ to be approximately constant during this motion, we can solve for the elapsed time using $v_{\mathrm{bary}}$ along with the distance between true ingress and approximate ingress:
\begin{equation*}
t_{\mathrm{cor,i}}=
\begin{cases}
    \frac{-\sqrt{(1+p)^2-b^2} - \left(-\sqrt{(1+R_{\mathrm{bin}}(t_i))^2-b^2}+\widetilde{a}_p\sin(\lambda_{sp})\right)}{v_{\mathrm{bary}}} & \mathrm{cases\;I, II} \\
    \frac{-\sqrt{(1+s)^2-b^2} - \left(-\sqrt{(1+R_{\mathrm{bin}}(t_i))^2-b^2}-\widetilde{a}_s\sin(\lambda_{sp})\right)}{v_{\mathrm{bary}}} & \mathrm{cases\;III, IV}
\end{cases}
\end{equation*}
For an illustrative reference, see Fig.~\ref{fig:impactcorr}. A similar treatment of the end-of-transit case gives:\\
\begin{equation*}t_{\mathrm{cor,e}}=
\begin{cases}
    \frac{\sqrt{(1+R_{\mathrm{bin}}(t_e))^2-b^2}+\widetilde{a}_p\sin\left(\lambda_{sp} + 2\pi\frac{T'_{14}}{P_s}\right) - \sqrt{(1+p)^2-b^2}} {v_{\mathrm{bary}}} & \mathrm{cases\;I, III} \\
    \frac{\sqrt{(1+R_{\mathrm{bin}}(t_e))^2-b^2}-\widetilde{a}_s\sin\left(\lambda_{sp} + 2\pi\frac{T'_{14}}{P_s}\right) - \sqrt{(1+s)^2-b^2}} {v_{\mathrm{bary}}} & \mathrm{cases\;II, IV}
\end{cases}
\end{equation*}
Then, under the approximation of circular orbits, we can make the substitution of
\begin{equation*}
v_{\mathrm{bary}} = \frac{2\pi \tilde{a}}{P}
\end{equation*}
The full approximation for small $b$ is then given by subtracting the appropriate $t_{\mathrm{cor,i}}$ amd $t_{\mathrm{cor,e}}$ from the corresponding $T'_{14}$ in Eq.~\ref{eq:A1}, Taylor approximated to $\sin^{-1}(\frac{l_1}{\tilde{a}}) + \sin^{-1}(\frac{l_2}{\tilde{a}}) \approx \frac{l_1+l_2}{\tilde{a}}$, which leads to further cancellation of several terms, ultimately resulting in Eq.~\ref{eq:2}.

\section{TTV-TDV behavior for sample binary planet systems}
As discussed in Sec.~\ref{sec:timedomain}.1, there is a significant short-period TTV signal associated with binary planet transits. In addition to the sample systems considered in Fig.~\ref{fig:ttv}, we compute the TTV-TDV behavior for a wider range of systems and show them in Fig.~\ref{fig:ttv-tdv_gallery} below. This exotic TTV-TDV behavior provides another observational feature to distinguish binary planets compared to low-mass planet companions.

\begin{figure}
    \centering
    \includegraphics[width=\textwidth]{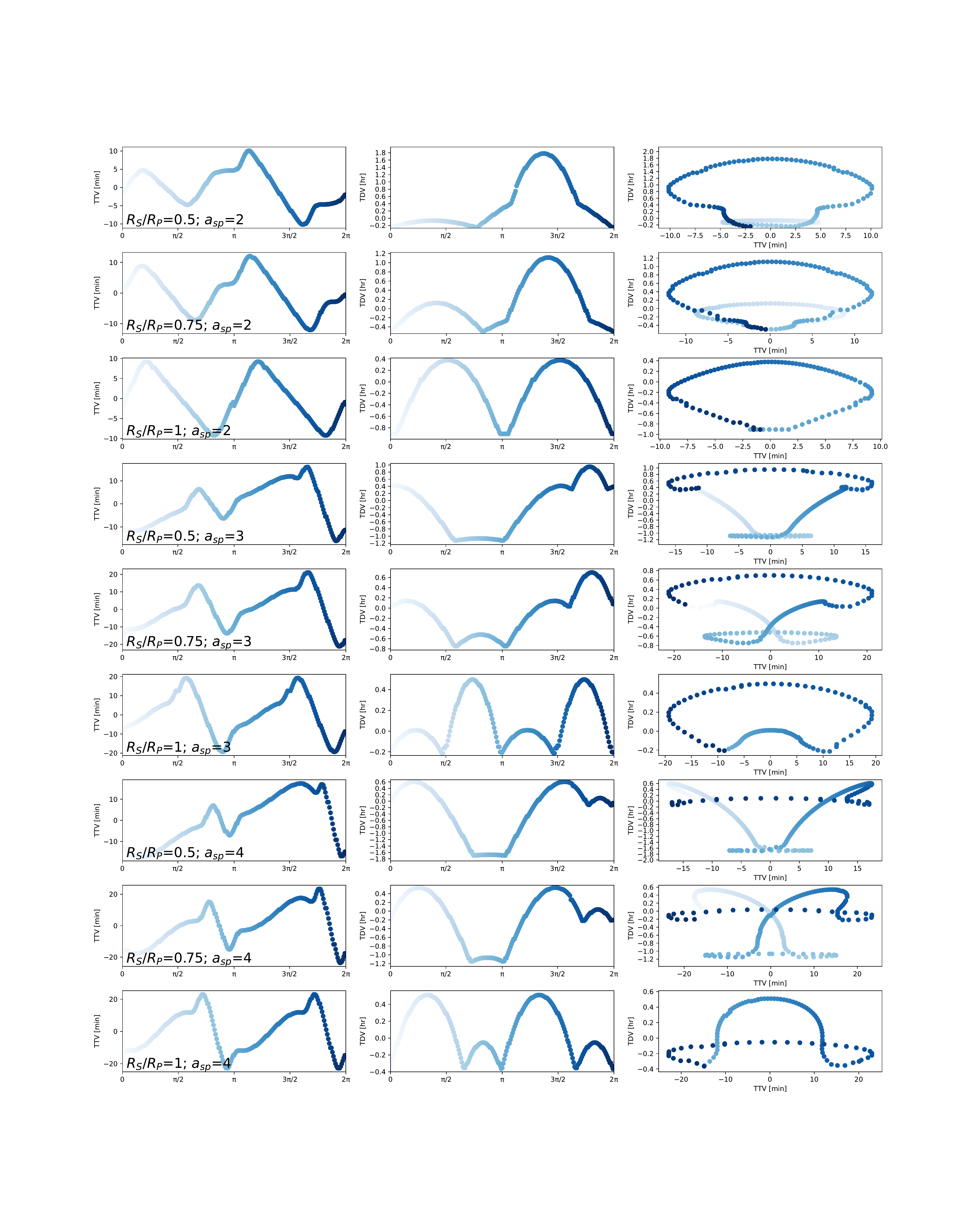}
    \caption{TTV, TDV, and joint plots for 9 sample binary planet systems across various values of $s$, $p$, and $a_{sp}$ (assuming Jupiter density for all planets). Opacity increases with $\lambda_{sp} \in [0, 2\pi)$.}
    \label{fig:ttv-tdv_gallery}
\end{figure}

\bsp
\label{lastpage}
\end{document}